\documentclass[preprint,12pt]{elsarticle}
\usepackage{CJK}
\usepackage{amsmath}
\usepackage{amssymb}
\usepackage[pdftex,colorlinks]{hyperref}
\usepackage{multirow}

\biboptions{compress}

\begin{document}
\begin{CJK*}{GBK}{song}
\begin{frontmatter}

\title{On the post-quantum security of encrypted key exchange protocols}

\author{Li Yang\corref{1}}
\author{Rui-Rui Zhou}
\cortext[1]{Corresponding author email: yangli@iie.ac.cn}
\address{State Key Laboratory of Information Security, Institute of Information Engineering, Chinese Academy of Sciences, Beijing 100093, China}
\begin{abstract}
We investigate the post-quantum security of the encrypted key exchange(EKE) protocols based on some basic physical parameters of ion-trap quantum computer, and show that the EKE protocol with a 40-bit password will be secure against a quantum adversary with several ion-trap quantum computers. We present a password encrypted no-key protocol to resist middle-man attack, and prove that it is also with the post-quantum security. The analysis presented here is probably of general meaning for the security evaluation of various hybrid cryptosystems.

\end{abstract}


\end{frontmatter}


The development of quantum computer is threatening modern cryptography, especially the public-key cryptosystems. The post-quantum security of cryptographic protocols becomes of general interest. Since there is no efficient quantum attack so far to the cryptographic algorithms based on hard problems of the lattice, coding and multivariate objects, many researchers turning their interest to related cryptography and cryptanalysis problem. Here we consider to resist this threat along another line: since a quantum computer is a physical system, its gate operation rate is limited by some basic physics parameters, thus some kinds of cryptographic algorithms may keep their security against a quantum adversary. We shall explain this through analyzing several examples in details.

\section{The basic idea}
The encrypted key exchange (EKE) protocol integrates the asymmetric and symmetric encryption algorithms to ensure two parties sharing a secret key$^{[1]}$. We first evaluate the post-quantum security of the EKE protocols via analyzing a simplified EKE protocol.
Suppose two participants $A(Alice)$ and $B(Bob)$ preshare a password $P$, they intend to establish a secret key $K$ via EKE protocol. A simplified EKE protocol is as follows$^{[2]}$:

Let $E_P(\cdot)/D_P(\cdot)$ be the encryption/decryption algorithms using password $P$ as its key, the public domain parameters are group $(G,\cdot)$, and $\alpha\in G$ with order $n$.

1. $A$ randomly chooses $\mu_A(0< \mu_A\leqslant n-1)$, and computes
\begin{equation}
\nu_A=\alpha^{\mu_A},y_A=E_P(\nu_A),
\end{equation}
then she sends $ID(A)$ and $y_A$ to $B$.

2. $B$ randomly chooses $\mu_B(0< \mu_B\leqslant n-1)$, and computes
\begin{equation}
\nu_B=\alpha^{\mu_B},y_B=E_P(\nu_B),
\end{equation}
then she sends $ID(B)$ and $y_B$ to $A$.

3. $A$ computes
\begin{equation}
\nu_B=D_P(y_B), K=(\nu_B)^{\mu_A}=\alpha^{\mu_B\cdot\mu_A}
\end{equation}
and $B$ computes
\begin{equation}
\nu_A=D_P(y_A), K={(\mu_A)}^{\nu_B}=\alpha^{\mu_A\cdot\mu_B}.
\end{equation}
Finally, $A$ and $B$ establish a key string $K$ with EKE protocol.

For an EKE protocol, the attacker(Eve) can get all the ciphertexts transmitted in the channel. As the password $P$ is not available to Eve, she cannot impersonate $B$ to interact with $A$ and get the session key $K$. In order to get $K$, Eve has to perform a password-guessing attack as follows:

She generates a candidate password $P^\prime$ to decrypt $E_P(\nu_A)$, $E_P(\nu_B)$, and obtains
\begin{equation}
{\mu_A}'=\log_{\alpha}\!{({\alpha}^{\mu_A})'},\ K^{\prime}=(\nu_B')^{{\mu_A}'},
\end{equation}
or
\begin{equation}
{\mu_B}'=\log_{\alpha}\!{({\alpha}^{\mu_B})'},\ K^{\prime}=(\nu_A')^{{\mu_B}'}.
\end{equation}

During this stage, Eve cannot check whether her candidate session key $K'$ is correct. She can only assess a $K'$ is correct by checking a decrypted plaintext is meaningful.

Suppose Eve has a quantum computer. For each candidate $P'$, Eve has to execute the discrete logarithm computation once. Let $\Delta T_1$ be a lower bound of time cost by a cycle of the discrete logarithm computation, and the time to perform an elementary quantum logic operation is $\Delta t_1$. It is well known that $\Delta t_1$ has a lower bound:
\begin{equation}
\Delta t_1\geqslant 10^{-14},
\end{equation}
then the time of a discrete logarithm computing cycle $\Delta T_1$ will be
\begin{equation}
\Delta T_1=N_1\cdot\Delta t_1,
\end{equation}
where $N_1$ is the number of quantum gate executed serially in a discrete logarithm algorithm. It can be seen that the value of $N_1$ depends on the improvement of paralization of quantum discrete logarithm algorithm. Although we cannot calculate the exact value of $N_1$, a rough estimation for a lower bound of $N_1$ is easy, then we have
\begin{equation}
\Delta T_1\geqslant 10^{4}\cdot\Delta t_1\geqslant 10^{4}\cdot10^{-14}=10^{-10},
\end{equation}
that is, the cycle of a discrete logarithm computation is no less than $10^{-10}$ seconds.

The physical security of a protocol$^{[4]}$ indicates that the protocol will never be broken by an adversary living in the real physical world. Let $N$ be the ultimate number of times a quantum computer can finish the discrete logarithm algorithm during 100 years, we have
\begin{equation}
N<\frac{2^{32}}{10^{-10}}< 2^{66}.
\end{equation}
For each candidate $P'$, Eve has to execute the quantum algorithm of discrete logarithm at least once to get $\mu_A'$ or $\mu_B'$, so the length of the password $P$ should satisfy $x\geqslant 66$, that is, for resisting attack with several($2^1$) quantum computers, a 68-bit password is enough to ensure the security of an EKE protocol within 100 years.

Above analysis is based on some universal parameters of single-qubit quantum gate operations. Now we focus on the analysis of CNOT gate of ion-trap quantum computers. Ion-trap quantum computer is proposed by Cirac and Zoller in 1995$^{[5]}$, it is considered one of the most promising scheme of physical implementation of quantum computer. To compute a discrete logarithm meaningful in cryptanalysis, the quantum computer must at least have
$n\sim2^{10}-2^{14}$ qubits. In an ion-trap quantum computer, the distance between two neighboring qubits is $a\sim$ 10 $\mu$m, and the characteristic distance of two qubits executing the CNOT operation is about
$10^{2}\ a$.
In an ion-trap, the mutual Coulomb force between two ions is
\begin{equation}
\frac{(n_0 e)^2}{4\pi\varepsilon_0\cdot r^2},
\end{equation}
where $n_0$ denotes the number of the electrons in an ion, and $r$ denotes the distance between the two ions. In the ion-trap quantum computer, the force to an ion can be expressed as
\begin{equation}
F=-K\cdot x,
\end{equation}
\begin{equation}
K=\frac{4(n_0 e)^2}{4\pi\varepsilon_0}(\frac{1}{1^3}+\frac{1}{2^3}+\cdots)\frac{1}{a^3}=\frac{4(n_0 e)^2}{4\pi\varepsilon_0}\zeta(3)\cdot\frac{1}{a^3},
\end{equation}
where $\zeta(\cdot)$ denotes the Riemann zeta-function, $\zeta(3)\approx1.2$, so
\begin{equation}
K=\frac{4.8(n_0 e)^2}{4\pi\varepsilon_0}\cdot\frac{1}{a^3}.
\end{equation}
It is proper to reckon the propagation velocity of interaction between different qubits carried by photons via evaluating the velocity of the acoustic waves in the trap based partially on some parameters of the stationary wave. For the ion-trap, the dispersion relation of the acoustic waves is
\begin{equation}
2\pi\nu=2\sqrt{\frac{K}{m}}\text{sin}\frac{\pi a}{\lambda},
\end{equation}
where $\nu,\lambda$ are the frequency and the wavelength of the acoustic wave, respectively, and $m$ is the mass of every ion. As $\lambda\gg a$, equivalent to  monochromatic approximation, the dispersion relation can be simplified as
\begin{equation}
2\pi\nu\approx2\sqrt{\frac{K}{m}}\frac{\pi a}{\lambda},
\end{equation}
then we obtain the velocity of the acoustic waves:
\begin{equation}
v=\lambda\cdot\nu=\sqrt{\frac{K}{m}}\cdot a=\sqrt{\frac{1.2(n_0 e)^2}{\pi\varepsilon_0am}}.
\end{equation}
Let $n_0=1$, the velocity will be
\begin{equation}
v=\sqrt{\frac{1.2e^2}{\pi\varepsilon_0am}}=0.32\times10^2 \ \ \text{m/s},
\end{equation}
so the time a CNOT operation takes will be greater than
\begin{equation}
\Delta t_2\approx2.85\times10^{-4}.
\end{equation}
In a discrete logarithm computation considered, the number of CNOT operations executed serially is $N_{2}\sim O(\text{log}\ \!n)^{[9]}$. In view of the fault-tolerant structure of the ion-trap quantum computer,especially the concatenated quantum error-correcting coding related with threshold theorem, we know that
\begin{equation}
N_2>10^2,
\end{equation}
then we can infer that the lower bound of time finishing a discrete logarithm computation is
\begin{equation}
\Delta T_2=N_2\cdot\Delta t_2>2.85\times10^{-4}\times10^2=2.85\times10^{-2}.
\end{equation}
That is, during one second, an ion-trap quantum computer cannot finish the discrete logarithm computation $2^6$ times, then the attacker cannot execute the discrete logarithm computation $2^{38}$ times in 100 years($\approx2^{32}$ seconds). Therefore, in 100 years, an EKE protocol with a 40-bit password can resist the attacker with several($2^1$) ion-trap quantum computers. Although physical parameters of other kinds of quantum computers are different from that of the ion-trap quantum computer, the conclusion is similar.

Let us estimate the length of a password that ensures the physical security of the EKE protocol.
Suppose the size of a quantum computer is about one meter, then the upper bound of number of the quantum computers employing by any adversary is
\begin{equation}
4\pi\times{(6370\times10^3)}^2=5.1\times10^{14}<2^{49}.
\end{equation}
With all these quantum computers the attacker can run the discrete logarithm algorithm about $2^{49+38}=2^{87}$ times within 100 years. When the length of the password $P$ is $88$, the attacker has to call the algorithm for averagely $2^{87}$ times, this is beyond the attacker's maximal calculating ability. 88 is roughly the number of bits of 11 characters of ASC$\uppercase\expandafter{\romannumeral2}$, which is an usual option of a password in nowadays internet. Our conclusion is: the attacker cannot break EKE protocol during 100 years even if he owns a huge number of ion-trap quantum computers, an 88-bit password is enough to achieve the physical security of a EKE protocol.
\section{Post-quantum security of several EKE protocols}
The EKE schemes use both symmetric and public key encryption algorithm to provide secrecy and entity authentication for the users of network$^{[10]}$. By executing an EKE protocol, two participants $A(Alice)$ and $B(Bob)$ can identify each other and generate a common key $K$.
\subsection{The EKE protocol}
The EKE protocol is as follows:

1. $A$ randomly generates the public key/secret key pair, $e_A$ and $d_A$. Then he uses the password $P$ to encrypt $e_A$ by symmetric encryption and sends to $B$
\begin{equation*}
A,E_P(e_A).
\end{equation*}

2. As $B$ is aware of the password $P$, he decrypts $E_P(e_A)$ to extract $e_A$. Afterwards, $B$ randomly generates a session key $K$, and encrypts it with the public key $e_A$ and the password $P$ respectively by asymmetric and symmetric encryption, then sends to $A$
\begin{equation*}
E_P(E_{e_A}(K)).
\end{equation*}

3. After receiving $E_P(E_{e_A}(K))$, $A$ uses the password $P$ and the secret key $d_A$ to extract $K$, then generates a random string $R_A$ and uses $K$ to encrypt it, then she sends to $B$
\begin{equation*}
E_K(R_A).
\end{equation*}

4. $B$ extracts $R_A$ with $K$, then generates another random string $R_B$, and sends to $A$
\begin{equation*}
E_K(R_A,R_B).
\end{equation*}

5. $A$ decrypts $E_K(R_A,R_B)$ and gets $(R_A, R_B)$, then she checks whether $R_A$ is correct. If the check succeeds, $A$ sends to $B$
\begin{equation*}
E_K(R_B).
\end{equation*}

6. $B$ decrypts $E_K(R_B)$ to get $R_B$, then he checks whether $R_B$ is correct. If the check pass, the protocol is accomplished, and $A$ and $B$ can use $K$ as the session key for communication.

An EKE protocol can be achieved with various public key algorithm. The one based on Diffie-Hellman protocol can be described as follows(the public domain parameters are group $(G,\cdot)$ and $g\in G$ with order $q$ which are public to all the users):

1. $A$ randomly chooses a random string $r_A$ and sends to $B$
\begin{equation}
A,g^{r_A}\ (\text{mod}\ q),
\end{equation}

2. $B$ randomly chooses a random string $r_B$ and computes: $K=g^{r_A\cdot r_B}$. He randomly chooses a string $R_B$, then computes and sends to $A$
\begin{equation}
E_P(g^{r_B} (\text{mod}\ q)), E_K(R_B).
\end{equation}

3. $A$ decrypts $E_P(g^{r_B} (\text{mod}\ q))$ to get $g^{r_B} (\text{mod}\ q)$, further she computes $K$ and uses it to extract $R_B$ by decrypting $E_K(R_B)$ with $K$. She generates another random string $R_A$, and sends to $B$
\begin{equation}
E_K(R_A,R_B).
\end{equation}

4. $B$ decrypts $E_K(R_A,R_B)$ to get $R_A$ and $R_B$. Once he ensures that the string $R_B$ is correct, he sends to $A$
\begin{equation}
E_K(R_A).
\end{equation}

5. $A$ decrypts $E_K(R_A)$ to get $R_A$. Once she ensures that the string $R_A$ is correct, the protocol is completed. Now the two parties can use $K$ as the session key for communication.

In this Diffie-Hellman-based EKE protocol, the attacker can get all the ciphertexts transmitted in the channel: $g^{r_A}(\text{mod}\ q)$, $E_P(g^{r_B} (\text{mod}\ q))$, $E_K(R_B)$, $E_K(R_A)$, and $E_K(R_A,R_B)$. After generating a candidate password $P'$, Eve can computes
\begin{equation}
r_B'=\log_g\!{[D_{P'}(E_P(g^{r_B} (\text{mod}\ q)))]},
\end{equation}
and further computes a candidate session key
\begin{equation}
K'=g^{r_A\cdot r_B'}.
\end{equation}
In order to avoid the attacker's achieving $g^{r_A}(\text{mod}\ q)$ directly, $A$ can send the encrypted string $E_P(g^{r_A}(\text{mod}\ q))$ to $B$. Then, the ciphertexts Eve obtains turn to be $E_P(g^{r_A}(\text{mod}\ q))$, $E_P(g^{r_B} (\text{mod}\ q))$, $E_K(R_B)$, $E_K(R_A)$ and $E_K(R_A,R_B)$. After generating a candidate password $P'$, Eve can computes a candidate key $K'$ as
\begin{equation}
r_B'=\log_g\!{[D_{P'}(E_P(g^{r_B} (\text{mod}\ q)))]},\ K^{\prime}=(D_{P'}(E_P(g^{r_A} (\text{mod}\ q))))^{r_B'}
\end{equation}
or
\begin{equation}
r_A'=\log_g\!{[D_{P'}(E_P(g^{r_A} (\text{mod}\ q)))]},\ K^{\prime}=(D_{P'}(E_P(g^{r_B} (\text{mod}\ q))))^{r_A'}.
\end{equation}
After getting $K'$, Eve can use it to decrypt $E_K(R_B)$, $E_K(R_A)$, $E_K(R_A,R_B)$ and obtains
\begin{equation}
R_A'=D_{K'}(E_K(R_A)),\ R_B'=D_{K'}(E_K(R_A)),
\end{equation}
\begin{equation}
(R_A'',R_B'')=D_{K'}(E_K(R_A,R_B))
\end{equation}
then verifies $K'$ by checking whether $(R_A',R_B')$ is equal to $(R_A'',R_B'')$.

We can see that, in this Diffie-Hellman-based EKE protocol the entity authentication can help Eve to check whether her candidate key is correct, at the same time, it will not reduce the number of of times executing the discrete logarithm computation. In this EKE protocol, for each candidate password $P'$, the attacker has to execute the discrete logarithm computation once, no matter whether $g^{r_A}$ is encrypted with the password $P$. Therefore, a 40-bit password is enough to resist an attacker with several($2^1$) quantum computers for 100 years, and an 88-bit password is enough to achieve a physical security of this protocol.
\subsection{The enhanced EKE protocol}
In the Diffie-Hellman-based EKE protocol, the attacker Eve an get all the ciphertexts transmitted in the channel: $g^{r_A}(\text{mod}\ q)$, $E_P(g^{r_B} (\text{mod}\ q))$, $E_K(R_B)$, $E_K(R_A)$ and $E_K(R_A,R_B)$. Once Eve recovers some old session keys, she may use these messages to launch some attacks about the password $P$. An enhanced EKE protocol$^{[11,12]}$ can efficiently forbid the attacker from using the old session key $K$ to get any information about the password $P$.

In the enhanced EKE protocol, the string $K=g^{r_A\cdot r_B}$ is used for key exchange. We denote the final session key as $S$ in step 2 of the Diffie-Hellman-based EKE protocol, B generates another random string $S_B$ and sends to $A$
\begin{equation}
E_P(g^{r_B} (\text{mod}\ q)),\ E_K(R_B,S_B),
\end{equation}
then, in step 3, $A$ generates another random string $S_A$ and sends to $B$
\begin{equation}
E_K(R_A,R_B,S_A).
\end{equation}
Finally, both $A$ and $B$ can get the session key $S=S_A\oplus S_B$. This key can be used to encrypt messages between $A$ and $B$, and $K$ is just used for encrypting random strings.

We can see that in this enhanced EKE protocol, the attacker Eve cannot directly get any information about the password $P$ from the session key $S$ because $P$ is not directly used to encrypt any string that can be used directly to derive $S$. In order to identify the password $P$, Eve should get the correct encryption key $K$ at first. Suppose Eve has already got an old session key $S$, he generate a candidate key $K'$ to computes
\begin{equation}
(R_A',R_B',S_A')=D_{K'}E_K(R_A,R_B,S_A),
\end{equation}
\begin{equation}
(R_B',S_B')=D_{K'}E_K(R_B,S_B),
\end{equation}
\begin{equation}
S'=S_A'\oplus S_B',
\end{equation}
then checks whether $K'$ is correct by checking whether $S'$ is equal to $S$. After getting the correct $K$, she generates a candidate password $P'$, then computes
\begin{equation}
r_B'=\log_g\!{[D_{P'}E_P(g^{r_B} (\text{mod}\ q))]},\ K'=g^{r_A\cdot r_B'},
\end{equation}
now Eve can check whether $P'$ is correct by checking whether $K'$ is equal to $K$.

Compared with the simple Diffie-Hellman-based EKE protocol, in this enhanced EKE protocol, the security of the password $P$ will not only depends on the length of the password $P$, but also depends on the length of the key $K$, and $K$ can be much longer than $P$. While $K$ is large enough, Eve cannot use the old session key $S$ to extract information about the password $P$. In addition, in this protocol the attacker also has to execute the discrete logarithm computation at least once for each candidate $P'$, so the calculated amount of the attacker is not reduced, the physical security of the protocol is still ensured, a 40-bits password is long enough to resist the attacker with several($2^1$) ion-trap quantum computer with 100 years, and an 88-bits password is long enough to achieve the physical security of the protocol.
\subsection{The A-EKE protocol}
The original EKE protocol has a defect: both the two parties $A$ and $B$ should be aware of the password $P$ in advance. Actually, most of identification systems based on password just store the hash value of the password, rather than the password itself. In an augmented EKE(A-EKE) protocol$^{[13]}$, the one-way hash value of the password $P$ is used as the super password. A Diffie-Hellman-based A-EKE protocol is as follows:

1. $A$ computes the one-way hash value of the password $P$(denoted by $p$) and randomly chooses a random string $R_A$, then she sends to $B$
\begin{equation}
E_{p}(g^{R_A}\ \text{mod}\ q).
\end{equation}

2. $B$ chooses a random string $R_B$, and sends to $A$
\begin{equation}
E_{p}(g^{R_B}\ \text{mod}\ q).
\end{equation}

3. Both $A$ and $B$ compute the session key
\begin{equation}
K=g^{R_A\times R_B}.
\end{equation}
In order to confirm that she really owns the password $P$, $A$ sends to $B$
\begin{equation}
E_K(S_P(K)).
\end{equation}
As $B$ is aware of the strings $K$ and $p$, he can easily decrypt $E_K(S_P(K))$ and verify the signature. Only the one who has the password $P$ and the session key $K$ can generate the cipher $E_K(S_P(K))$. As analyzed in the simple EKE protocol in Sec.1, the attacker also has to run the discrete logarithm algorithm once for each candidate $p'$ in the A-EKE protocol, so the physical security of this protocol is ensured by an 88-bit password. It can be seen that the attacker cannot get any information about the password $P$.
\section{Post-quantum security of encrypted no-key protocol}
\subsection{Shamir's no-key protocol}
In addition to the EKE protocol, other hybrid cryptosystems are of interest for post-quantum cryptography, such as Shamir's no-key protocol. Shamir's no-key protocol can be regarded as a kind of key generation protocol$^{[14]}$. In this protocol, two participants $A$ and $B$ exchange three messages through a public channel without authentication to transmit a secret key. The basic no-key protocol is as follows(a prime number $q$ is a public parameter that chosen in advance):

1. $A$ randomly chooses a key string $K$ and a secret number $a$, then she computes $a^{-1}\ (\text{mod}\ q-1)$ and sends to $B$
\begin{equation}
K^a\ \text{mod}\ q
\end{equation}

2. $B$ randomly chooses a secret number $b$ and computes $b^{-1}\ (\text{mod}\ q-1)$, and sends to $A$
\begin{equation}
(K^a)^b\ \text{mod}\ q
\end{equation}

3. $A$ computes
\begin{equation}
K^b(\text{mod}\ q)=((K^a)^b)^{(a^{-1} \text{mod}\ q-1)} \text{mod}\ q
\end{equation}
and sends it to $B$.

4. $B$ recovers the secret key $K$ via computing
\begin{equation}
K=(K^b)^{(b^{-1}\ \text{mod}\ q-1)}\ \text{mod}\ q.
\end{equation}
Then, $A$ transmits the secret key $K$ to $B$ successfully.

It can be seen that the basic no-key protocol cannot resist the middle-man attack. The attacker Eve can get $K$ as follows:

Firstly, Eve impersonates $B$ to interact with $A$:
\begin{gather*}
1. A\rightarrow Eve:K^a\ \text{mod}\ q,\\
\ 2. A\leftarrow Eve:K^{ae}\ \text{mod}\ q,\\
3. A\rightarrow Eve:K^e\ \text{mod}\ q.
\end{gather*}
Then, Eve computes the secret key $K$ and personates $A$ to interact with $B$:
\begin{gather*}
1. Eve\rightarrow B:K^e\ \text{mod}\ q,\\
\ 2. Eve\leftarrow B:K^{eb}\ \text{mod}\ q,\\
3. Eve\rightarrow B:K^b\ \text{mod}\ q.
\end{gather*}

\subsection{The encrypted no-key protocol}
In order to resist the middle-man attack, we use a password shared by $A$ and $B$ to enhance the no-key protocol.

The protocol is as follows:

1. $A$ randomly generates the secret key $K$ and a random number $a$, and computes $K^a\ \text{mod}\ q$, then uses the password $P$ to encrypt it by a given symmetric encryption algorithm and sends to $B$
\begin{equation}
E_P(K^a\ \text{mod}\ q).
\end{equation}

2. $B$ decrypts $E_P(K^a\  \text{mod}\ q)$ and then randomly generates a string $b$. He computes $K^{ab}\ \text{mod}\ q$, and
encrypts it with the password $P$, and then sends to $A$
\begin{equation}
E_P(K^{ab}\ \text{mod}\ q).
\end{equation}

3. After receiving $E_P(K^{ab}\ \text{mod}\ q)$, $A$ uses the password $P$ to extract $K^{ab}\ \text{mod}\ q$, then she computes
\begin{equation}
K^b(\text{mod}\ q)=(K^{ab})^{a^{-1}}\ \text{mod}\ q,
\end{equation}
and sends to $B$
\begin{equation}
E_P(K^b\ \text{mod}\ q).
\end{equation}

4. $B$ decrypts $E_P(K^b\ \text{mod}\ q)$ to get the secret key $K$,
\begin{equation}
K=D_P(E_P(K^b\ \text{mod}\ q))^{b^{-1}}\ \text{mod}\ q.
\end{equation}


For this encrypted no-key protocol, Eve can get all the ciphertexts transmitted in the channel: $E_P(K^a\ \text{mod}\ q)$, $E_P(K^{ab}\ \text{mod}\ q)$, $E_P(K^b\ \text{mod}\ q)$. As password $P$ is not available, Eve cannot get $K$ via impersonating $B$ to interact with $A$. In order to obtain $K$, she has to perform a password-guessing attack as follows:

Eve randomly generates a candidate password $P^\prime$, then uses it to decrypt $E_P(K^a\ \text{mod}\ q)$, $E_P(K^{ab}\ \text{mod}\ q)$ and $E_P(K^b\ \text{mod}\ q)$, and obtains $(K^a)'$, $(K^b)'$ and $(K^{ab})'$. Then, she try to extract $a'$ from $[(K^b)',(K^{ab})']$, and $b'$ from $[(K^a)',(K^{ab})']$. Finally, she computes
\begin{equation}
 K'=((K^a)')^{(a')^{-1}},
\end{equation}
\begin{equation}
K''=((K^b)')^{(b')^{-1}},
\end{equation}
and verifies whether the candidate password $P'$ is correct by checking whether $K^\prime$ is equal to $K''$. It is obvious that the computational complexity of this attack depends on the length of password $P$.

For each candidate $P'$, Eve has to execute the discrete logarithm computation twice. As analyzed in Sec.1, in order to resist an attacker with an ion-trap quantum computer, the length $x$ of the password $P$ should satisfy:
\begin{equation}
2^x>2^{38}/2=2^{37},
\end{equation}
thus, for an attacker with several($2^2$) ion-trap quantum computers, a 40-bit password is enough to ensure the physical security of the protocol within 100 years. Account of the attacker's maximal ability of computing in physics, an 88-bit password is enough.

The hybrid cryptosystems integrating both asymmetric and symmetric algorithms have already been a common choice in the practice of cryptography. The problem is to evaluate their security in quantum computing environment, and to choose appropriate parameters in accordance with their security requirements. For the developing of post-quantum cryptography, this is in fact one of the most direct and effective option.

\end{CJK*}
\end{document}